\documentclass[final,5p]{elsarticle}
\cortext[cor1]{Corresponding author: \texttt{alexander.gregg@newcastle.edu.au}}
\journal{Nuclear Instruments and Methods in Physics Research Section B}
\biboptions{sort&compress}

\usepackage{amsmath,amssymb,amsfonts,amsthm,enumerate,multirow}
\usepackage{placeins}
\usepackage{dblfloatfix}
\usepackage[percent]{overpic}
\usepackage{enumitem}
\usepackage{xcolor}
\usepackage[utf8]{inputenc}

\newcommand{\bs}[1]{\boldsymbol{#1}}

\newcommand{\tL}{\boldsymbol{\mathcal{L}}}

\newcommand{\Transp}{\mathsf{T}}
\newcommand{\dd}{\; \text{d}}
\newcommand{\nhat}{\hat{\boldsymbol{n}}}
\newcommand{\bepsilon}{\boldsymbol{\epsilon}}

\newcommand{\kfold}{$k-$fold }
\newcommand{\norm}[1]{\left\lVert#1\right\rVert}
\begin{document}

\begin{frontmatter}
\title{Radial Basis Functions and Improved Hyperparameter Optimisation for Gaussian Process Strain Estimation}

\author[NewcE]{A.W.T. Gregg\corref{cor1}}
\author[NewcE]{J.N. Hendriks}
\author[NewcE]{C.M. Wensrich}
\author[NewcE]{N. O'Dell}
\address[NewcE]{School of Engineering, The University of Newcastle, Callaghan NSW 2308, Australia}

\begin{abstract}
Over the past decade, a number of algorithms for full-field elastic strain estimation from neutron and X-ray measurements have been published. Many of the recently published algorithms rely on modelling the unknown strain field as a Gaussian Process (GP) - a probabilistic machine-learning technique. Thus far, GP-based algorithms have assumed a high degree of smoothness and continuity in the unknown strain field.
In this paper, we propose three modifications to the GP approach to improve performance, primarily when this is not the case (e.g. for high-gradient or discontinuous fields); hyperparameter optimisation using \kfold cross-validation, a radial basis function approximation scheme, and gradient-based placement of these functions.
\end{abstract}
\end{frontmatter}

\section{Introduction}
Over the past decade, a number of algorithms for full-field elastic strain tensor reconstruction from neutron and X-ray measurements have been published \cite{korsunsky2006,korsunsky2011,abbey09,abbey12,kirkwood15,wensrich16a,wensrich16b,gregg2017axi,gregg2018resid,hendriks2018traction,jidling2018probabilistic,hendriks2019robust,kirkwood2019application,sato2014,hendriksxray,hendriks3d,greggdt}. For the most part, these can be broadly classified as solutions to `rich' tomography problems --- the reconstruction of higher-order tensor fields from lower-order (average) measurements.

Amongst these are algorithms that operate on Bragg-edge neutron transmission images \cite{abbey09,abbey12,kirkwood15,wensrich16a,wensrich16b,gregg2017axi,gregg2018resid,hendriks2018traction,jidling2018probabilistic,kirkwood2019application,sato2014,hendriks3d}, conventional diffraction strain scans \cite{hendriks2019robust}, high-energy X-ray measurements \cite{korsunsky2006,korsunsky2011,hendriksxray}, and most recently so-called `diffraction tomography' profiles \cite{greggdt}. The particular details of these measurements and algorithms are described in detail in the provided references. Literature reviews that place these in context and discuss the differences and relative benefits of each approach can be found in \cite{hendriksthesis}, \cite{gregg2018resid} and \cite{greggdt}.

Many of the recently published algrothims \cite{hendriks2018traction,jidling2018probabilistic,hendriks2019robust,hendriksxray,hendriks3d,greggdt} implement Gaussian Process (GP) regression. This machine learning technique is detailed in \cite{rasmussen2006gaussian}, and was first demonstrated in this context by \cite{jidling2018probabilistic}, which provides the framework for application of the GP method to strain estimation.

So far, GP-based approaches have assumed a high degree of smoothness and continuity in strain by their underlying choice of covariance function. In the vast majority of cases this assumption holds true, and consequently these approaches have successfully reconstructed a number of strain fields from both simulated and real-world measurements \cite{hendriks2018traction,jidling2018probabilistic,hendriks2019robust,hendriksxray,hendriks3d,greggdt}. Assuming smoothness is troublesome when reconstructing high-gradient or discontinuous strain fields (e.g. in shrink-fit samples or multi-body assemblies).

In this paper we propose three developments to the GP approach to improve performance, particularly in this case; \kfold cross-validation for hyperparameter optimisation, an alternative approximation scheme for the GP using \emph{Radial Basis Functions} (RBFs), and a gradient-based approach to RBF placement and refinement of these functions. We demonstrate these concepts for simple 1-D examples before comparing the modified approach to a previously published GP-based algorithm {\cite{hendriks2018traction}. This comparison is made using experimental Bragg-edge neutron transmission measurements of a discontinous 2D ring-and-plug residual strain field \cite{gregg2018resid}.

\section{A Brief Review of Gaussian Processes}
\label{sec:GPintro}
A detailed introduction to GP regression is provided in \citep{rasmussen2006gaussian}, and the specifics related to implementing this technique for strain reconstruction can be found in \citep{jidling2018probabilistic}.

Briefly, a GP models an unknown field as a Gaussian distribution of random functions $\bs{f(x)}, \; \bs{x} \in \mathbb{R}^{\text{dim}(\bs{x})}$, described by mean $\bs{m(x)}$ and covariance functions $\bs{K(x,x')}$, where;
\begin{align*}
\bs{m(x)} &= \mathbb{E}\left[\bs{f(x)}\right], \\
\bs{K(x,x')} &= \mathbb{E}\left[(\bs{f(x)}-\bs{m(x)})(\bs{f(x')}-\bs{m(x')})^\Transp\right].
\end{align*}

GP regression estimates a function value at a query point $\bs{x_*}$ from a set of data, $\mathcal{D} = \left\{y_i,\bs{\eta}_i \; \vert \; \forall \; i=1,\dots,n\right\}$, assuming each measurement is of the form;
\begin{equation*}
y_i = \tL_{\bs{\eta}_i} \bs{f(x)} + e_i,
\end{equation*}
where $\tL_{\bs{\eta}_i} \bs{f(x)}$ is a linear transformation of $\bs{f(x)}$, parametrised by the set $\bs{\eta}_i$, and the measurement noise $e_i\sim \mathcal{N}(0,\sigma_i^2)$ is assumed zero-mean and Gaussian with variance $\sigma_i^2$.

GPs are closed under linear operators \citep{papoulis2002probability,wahlstrom2015modeling}, meaning the measurements $\bs{Y} = [y_1, y_2, \ldots y_n]^\Transp$ and a function value estimate $\hat{f}(\bs{x_*})$ are jointly Gaussian \citep{rasmussen2006gaussian};
\begin{equation*}
\begin{bmatrix}
\bs{Y} \\
\bs{\hat{f}(x_*)}
\end{bmatrix} \sim \mathcal{N} \left(
\begin{bmatrix}
\bs{\mu_y} \\
\bs{m(x_*)}
\end{bmatrix},
\begin{bmatrix}
\bs{K_{yy'}+\Sigma_m} & \bs{K_{y\hat{f}'}}\\
\bs{K_{\hat{f}y'}} & \bs{K(x_*,x_*)}
\end{bmatrix} 
\right).
\end{equation*}

Above, $\bs{\Sigma_m}$ is a diagonal matrix with nonzero elements containing the measurement variances. The cross-covariance matrices $\bs{K_{y\hat{f}'}} = \bs{K_{\hat{f}y'}}^\Transp$ and covariance matrix $\bs{K_{yy'}}$ are given by:
\begin{equation*}
    \bs{K_{y\hat{f}_*}} = \begin{bmatrix}
        \tL_{\bs{\eta}_1} \bs{K}(\bs{x},\bs{x_*}) \\
        \vdots \\
        \tL_{\bs{\eta}_n} \bs{K}(\bs{x},\bs{x_*})
    \end{bmatrix} 
\end{equation*}
and
\begin{equation*}
\begingroup 
\setlength\arraycolsep{1pt}
    \bs{K}_{\bs{y}\bs{y}'} = \begin{bmatrix}
        \tL_{\bs{\eta}_1} \bs{K}(\bs{x},\bs{x'})\tL_{\bs{\eta}_1} '^\Transp & \cdots & \tL_{\bs{\eta}_1} \bs{K}(\bs{x},\bs{x'})\tL_{\bs{\eta}_n}'^\Transp  \\
         \vdots & \ddots & \vdots \\ 
         \tL_{\bs{\eta}_n} \bs{K}(\bs{x},\bs{x'})\tL_{\bs{\eta}_1} '^\Transp & \cdots & \tL_{\bs{\eta}_n} \bs{K}(\bs{x},\bs{x'})\tL_{\bs{\eta}_n} '^\Transp \\ 
    \end{bmatrix}.
\endgroup
\end{equation*}

With these covariances, the prior $f(\bs{x_*})$ can be conditioned on the measurements to give a posterior estimate with mean and variance according to the closed-form expressions;
\begin{align}
\begin{split}
        \bs{\mu}_{\bs{f}_*|\bs{Y}} &= \bs{m}(\bs{x}_*) + \bs{K_{\hat{f}y'}}\left(\bs{K}_{\bs{yy}'}+\bs{\Sigma_m} \right)^{-1}(\bs{Y}-\bs\mu_y), \\
        \Sigma_{\bs{f}_*|\bs{Y}}  &= \bs{K}(\bs{x}_*,\bs{x}_*) - \bs{K_{\hat{f}y'}}\left(\bs{K}_{\bs{yy}'}+\bs{\Sigma_m}\right)^{-1}\bs{K_{y\hat{f}'}}.
\end{split}
\label{eqn:closedform}
\end{align}

\section{Encoding Physical Constraints}
It is possible to encode physical constraints in the reconstruction process. For instance, application of equilibrium (either directly using the strong form or by a minimisation of strain energy --- the weak form) has been central to a number of previously published algorithms \cite{kirkwood15,wensrich16a,wensrich16b,gregg2017axi,gregg2018resid,hendriks2018traction,jidling2018probabilistic,hendriks2019robust,kirkwood2019application,hendriksxray,hendriks3d,greggdt}. In general, this constraint aids in convergence and provides a physically viable solution. Additionally, in the case of Bragg-edge neutron transmissions measurements, this constraint is necessary to provide a unique reconstruction given the nontrivial null space of the Longitudinal Ray Transform (LRT) measurement model {\cite{lionheart15,jidling2018probabilistic,hendriks2018traction}. 

Equilibrium has been encoded in prior works by constructing a GP for a potential function $\phi$ from which the strains $\boldsymbol{\epsilon}$ are derived {\cite{jidling2018probabilistic}. In 2D, this was achieved by an Airy Stress potential \cite{jidling2018probabilistic,hendriks2018traction,hendriks2019robust}, and in 3D using Beltrami stress functions \cite{hendriksxray,hendriks3d}.

Implementation of this constraint is not without cost. As mentioned in Section \ref{sec:GPintro}, any linear transformation $\tL$ of the underlying GP (e.g. from $\phi$ to $\boldsymbol{\epsilon}$, or from $\boldsymbol{\epsilon}$ to a measurement $y$) must be applied to the covariance matrices twice (i.e. once as $\tL$ and once as $\tL'^\Transp$). Not only does this require that these transformations are possible (e.g. a 2nd-derivative transformation requires that the fourth-derivative of the function exists), but implementation of these can introduce significant computational burden in the calculation of these matrices.

As an illustrative example, consider the case of 2D Bragg-edge neutron transmission measurements, as described in \cite{jidling2018probabilistic}. Here, the strains $\boldsymbol{\epsilon}$ are related to an Airy stress function potential $\phi$ according to the classical mapping and Hooke's law\footnote{Note: the expression provided here is for a plane-stress assumption. A plane-strain formulation differs slightly.} as follows:
\begin{equation*}
\boldsymbol{\epsilon} = \begin{bmatrix} \epsilon_{xx} \\ \epsilon_{xy} \\ \epsilon_{yy}\end{bmatrix} = \begin{bmatrix} \frac{\partial^2}{\partial y^2}- \nu\frac{\partial^2}{\partial x^2}\\ -(1+\nu)\frac{\partial^2}{\partial x \partial y} \\ \frac{\partial^2}{\partial x^2}- \nu\frac{\partial^2}{\partial y^2}\end{bmatrix}\phi = \tL_{\bs{x}}\phi,
\end{equation*}
where $\nu$ is Poisson's ratio. With respect to the sample geometry and coordinate system given \cite{gregg2018resid}, Bragg-edge neutron transmission measurements are modelled by the LRT: a line integral average of the normal component of strain seen by a ray from $s=0$ (where the ray enters the sample) to $s=L$ (where it leaves):
\begin{equation*}
y=\frac{1}{L}\int_0^L \nhat^\Transp \bepsilon (s) \nhat  \dd s = \tL_y \boldsymbol{\epsilon},
\end{equation*}

Accordingly, two transformations must be performed to relate Bragg-edge neutron transmission measurements to the underlying Airy Stress function potential for which we construct a GP:
\begin{equation*}
y = \tL_y \boldsymbol{\epsilon}= \tL_y \tL_{\bs{x}} \phi.
\end{equation*}

In order to calculate $\bs{K}_{\bs{y}\bs{y}'}$ for example, this mapping must be applied twice, and a double integral of fourth-derivatives of the covariance function must be evaluated:
\begin{align*}
\bs{K}_{\bs{y}\bs{y}'} &= \tL_y \tL_{\bs{x}} \bs{K}(\bs{x},\bs{x_*}) \tL_{\bs{x}}^\Transp \tL_y^\Transp \\
& = \int_0^{L_i}\int_0^{L_j} \begin{bmatrix} n_{i,x}^2 & 2n_{i,x}n_{i,x} & n_{i,y}^2 \end{bmatrix} \begin{bmatrix} \frac{\partial^2}{\partial y^2}- \nu\frac{\partial^2}{\partial x^2}\\ -(1+\nu)\frac{\partial^2}{\partial x \partial y}- \nu\frac{\partial^2}{\partial x^2} \\ \frac{\partial^2}{\partial x^2}- \nu\frac{\partial^2}{\partial y^2}\end{bmatrix} \\
& \qquad \qquad \bs{K}(\bs{x_{0,i}} + s_i \nhat_i,\bs{x_{0,j}} + s_j \nhat_j) \\
& \begin{bmatrix} \frac{\partial^2}{\partial y^2}- \nu\frac{\partial^2}{\partial x^2}\\ -(1+\nu)\frac{\partial^2}{\partial x \partial y}- \nu\frac{\partial^2}{\partial x^2} \\ \frac{\partial^2}{\partial x^2}- \nu\frac{\partial^2}{\partial y^2}\end{bmatrix}^\Transp \begin{bmatrix} n_{j,x}^2 & 2n_{j,x}n_{j,x} & n_{j,y}^2 \end{bmatrix}^\Transp  \dd s_i \dd s_j
\end{align*}

The same process is required for other measurement models, and is in some cases even more burdensome. For example, area integrals are used to model the average strain within a gauge volume in 2D for conventional `point-wise' diffraction measurements \cite{hendriks2019robust}. To apply the exact GP approach, this measurement model would again have to be applied twice, and four integrals of fourth-derivatives of the covariance function would be required.

Analytical derivatives and integrals of the covariance function can sometimes be calculated \cite{jidling2018probabilistic}, reducing this burden significantly.
In other cases, integrals and derivatives must be calculated numerically \cite{greggdt}, --- a time-consuming and potentially unstable approach when applied to large data sets (e.g. Bragg-edge strain tomography, where 25,000+ measurements can been seen \cite{hendriks2018traction,hendriks2017,gregg2018resid} depending on the binning of neutron counts).

This problem is not insurmountable --- combinations of these approaches (e.g. an analytical solution to a first integral, followed by a numerical approach to a second) have shown some promise \cite{hendriks2018traction}, and approximation schemes, discussed in Section \ref{sec:approx}, have proven to be a convenient and robust means of simplifying this process, even for large data sets.

Nevertheless, careful consideration must be paid to the transformations required to both encode physical constraints and implement measurement models when selecting a covariance function.

\section{Covariance Functions and Hyperparameter Optimisation}

Selection of a covariance function $K(x,x')$ and the associated hyperparameters can a have a profound impact on the resulting reconstruction.

Thus far, most GP-based approaches to strain tomography have implemented the explicit form of, or an approximation to the squared exponential kernel:
\begin{equation*}
\bs{K(x,x')} = \sigma_f^2 \; \text{exp}\left(\frac{-\norm{\bs{x}-\bs{x'}}^2}{2\ell^2}\right),
\end{equation*}
which is characteried by the hyperparameters $\sigma_f^2$ (a prior variance) and $\ell$, a length-scale.

This kernel assumes a high degree of smoothness, and in many cases has proven to be a good choice in modelling strain --- an ordinarily smooth phenomena.  

The hyperparameters of this covariance function have thus far been tuned using the measurements (i.e. with no a-priori knowlege) by a marginal likelihood maximisation routine. This optimsation places costs on both suitability of the hyperparameters according to the measurements and on model complexity - limiting over-fit \cite{jidling2018probabilistic}.

With that said, high-gradient and/or discontinous strain fields present two significant challenges for these approaches:
\begin{enumerate}
    \item These fields break the fundamental modelling assumption of smoothness intrinsic to this choice of kernel.
    \item The weighting given to the zero-mean prior and penalty on complexity intrinsic to the marginal likelihood maximisation process can inhibit selection of small-enough length-scales to capture regions containing high gradients or discontinuities.
\end{enumerate}

\section{Approximation Schemes and Basis Functions}
\label{sec:approx}

To both simplify the implementation of physical constraints and consequently reduce the computational burden associated with large data sets, many of the published GP reconstruction algorithms have used an approximation scheme to represent $\bs{K(x,x')}$ with a finite sum of basis functions \cite{jidling2018probabilistic}, according to:
\begin{equation}
     K_\varphi(\mathbf{x},\mathbf{x}') = \Phi(\mathbf{x})\Sigma_p\Phi(\mathbf{x}')^\top,
\end{equation}
where each column of $\Phi(\mathbf{x})$ is a basis function $\phi_j(\mathbf{x})$ with spectral density $\Sigma_{p,jj}$.

This formulation only requires that linear transformations (such as that encoding equilibrium or implementing a measurement model) be applied once to the basis function, rather than twice to the covariance function, in-general greatly simplifying this process. The specifics of implementing this scheme for strain reconstruction are detailed in \cite{jidling2018probabilistic} and elaborated on in \cite{hendriks2019robust}.

When using an approximation scheme, instead of forming the full joint prior distribution and reconstructing using Equation \ref{eqn:closedform}, an estimate of $f_*$ is instead calculated by:
\begin{equation*}
    \begin{split}
        \boldsymbol\mu_{\boldsymbol f_* | \mathbf{Y}} &= \Phi_*\left(\Phi_Y(\mathbf{x})^\top\Sigma_Y^{-1}\Phi_Y(\mathbf{x})+\Sigma_p^{-1}\right)^{-1}\Phi_Y(\mathbf{x})^\top\Sigma_Y^{-1}(\mathbf{Y}-\boldsymbol\mu_y) \\
         \boldsymbol\Sigma_{\boldsymbol f_* | \mathbf{Y}} &= \Phi_*\left(\Phi_Y(\mathbf{x})^\top\Sigma_Y^{-1}\Phi(\mathbf{x})+\Sigma_p^{-1}\right)^{-1}\Phi_*^\top.
    \end{split}
\end{equation*}

A numerically robust approach to these calculations is provided in \cite{hendriks2019robust}, and involves the use of the QR decomposition to compute the required matrix inverses.

Prior approaches \cite{jidling2018probabilistic,hendriks2019robust,hendriksxray,hendriks3d} have utilised a harmonic approximation to the squared-exponential covariance function:
\begin{equation}\label{RI:eq:scalar_baseis}
\begin{split}
    \phi_j(\mathbf{x}) &= \frac{1}{\sqrt{L_xL_y}}\sin(\lambda_{xj}(x+L_x))\sin(\lambda_{yj}(y+L_y)), \\
    \Sigma_{p,jj} &= \sigma_f^2 2\pi l_x l_y \exp\left(-\frac{1}{2}\left(l_x^2\lambda_{xj}^2+l_y^2\lambda_{yj}^2\right)\right).
\end{split}
\end{equation}

where $L_x, L_y, \lambda_{xj}$ and $\lambda_{yj}$ control the frequency and phase of the basis functions. These quantities are chosen by a constrainted hyperparameter optimisation process such that the basis functions span a region where their spectral densities, $\Sigma_{pjj}$, are greater than a minimum threshold, helping to ensure that the dominant frequencies of the response are captured while maintaining numerical stability.

Note that this approach bears some similarity to \cite{gregg2018resid}, which could be viewed as giving the maximium likelihood solution (least-squares), though the major limitations of that algorithm are addressed intrinsically by using a GP:

\begin{enumerate}
    \item The frequencies of the basis functions are selected automatically from the measurements by the hyperparameter optimisation process.
    \item The equilibrium constraint is automatically encoded in the solution, and is applied universally throughout the field.
    \item A closed-form for the mean and variance of the reconstruction exists - no optimisation is required to `fit' the basis functions.
\end{enumerate}

The harmonic approximation scheme has proven successful in a range of circumstances and by nature provides good flexibility with relatively few basis functions. That being said, these basis functions still present a number of issues when reconstructing high gradient or discontinous strain fields:

\begin{enumerate}
    \item Prior approaches still approximate stationary, smooth covariance functions \cite{jidling2018probabilistic,hendriks2018traction,hendriks2019robust,hendriks3d}, and a fundamental model mis-match is still present.
    \item The periodic and ongoing nature of harmonic basis functions means that misfit (e.g. due to a discontinuity) tends to progate to the rest of the reconstruction (`ringing' artefacts, as shown in Figure \ref{fig:fig3}).
    \item The choice of which frequencies to include is not obvious --- complicating the process of constraining the hyperparameter optimisation, or determining how many basis functions to include. Selective refinement around high-gradient or discontinuous features is also not possible.
\end{enumerate}

\section{Proposed Developments to the GP Technique}

In this section we propose three developments to the GP-based strain reconstruction approach; $k-$fold Cross Validation for hyperparameter optimisation, the use of Radial Basis Functions (RBFs) in an approximation scheme, and gradient-based RBF Placement. For simplicity, these improvements are first discussed in the context of, and demonstrated for the 1D example shown in Figure \ref{fig:fig1}.

The underlying function being estimated is a shifted unit step $f(x) = \mu(x-0.5)$, and reconstructions are made from 200 equally spaced point measurements on the domain $x\in \begin{bmatrix} 0 & 1 \end{bmatrix}$, corrupted by mean zero simulated gaussian noise with standard deviation $\sigma_n = 0.05$.

Following these 1D demonstrations, the cumulative effect of the proposed developments are explored on experimental data in Section \ref{sec:exp}.

\begin{figure}[h!]
\begin{center}
    \includegraphics[width=0.8\linewidth]{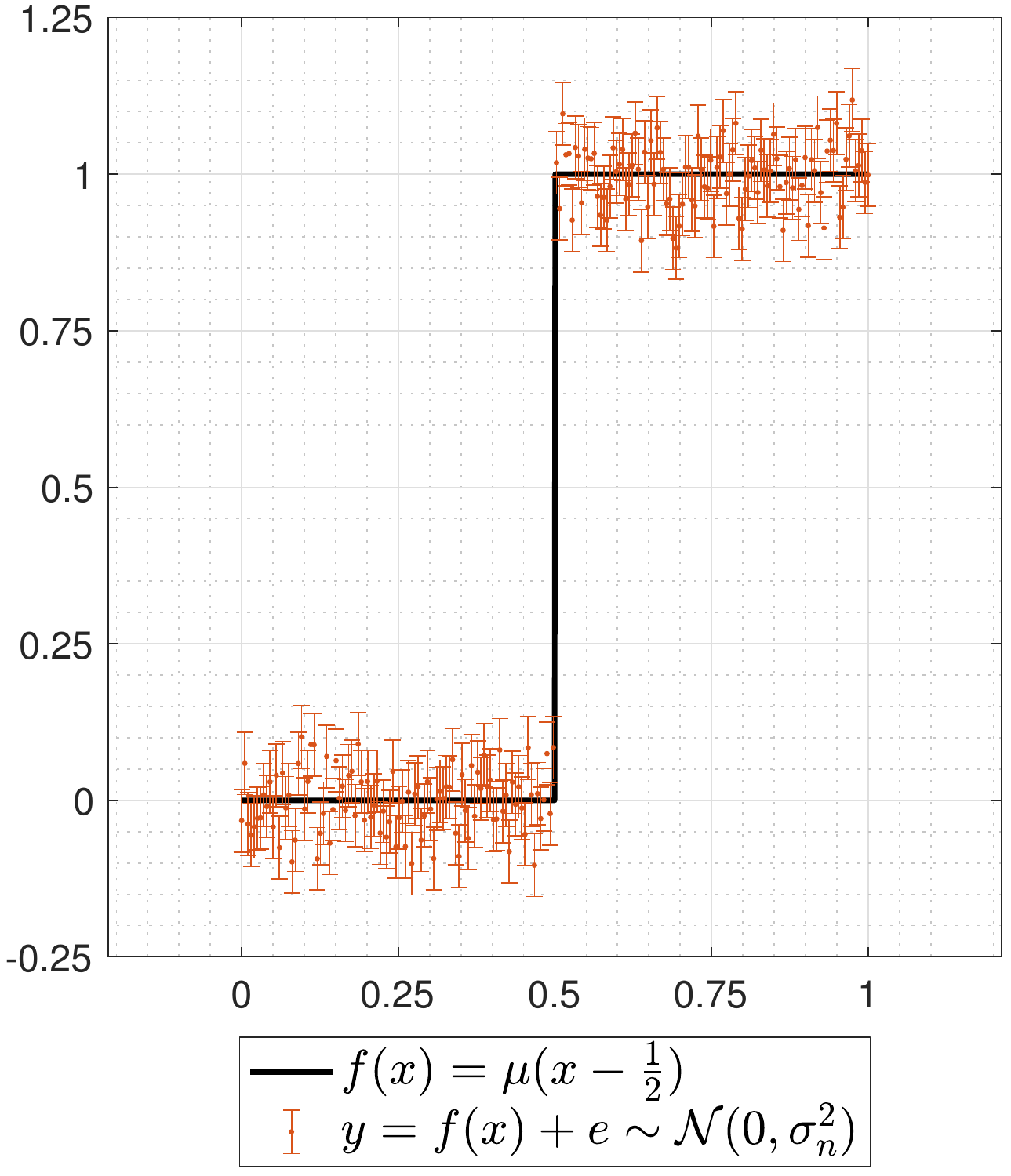}
\caption{Example unit step function and measurements.}
    \label{fig:fig1}
\end{center}
\end{figure} 

\FloatBarrier

\FloatBarrier
\subsection{\kfold Cross Validation for Hyperparameter Optimisation}
\label{subsec:crossval}

\kfold Cross Validation is an alternative approach to hyperparameter optimisation that can improve performance in the case of kernel mis-specification \cite{rasmussen2006gaussian,wabba}, such as when estimating a discontinuous field with a smooth model.

The process of implementing this alternative approach is discussed at length in \cite{rasmussen2006gaussian}, but a brief outline is as follows:

\begin{enumerate}
    \item The measured observations are randomly divided into two bins - a training set, and a validation set. The ratio of divided data is typically such the training set is much larger than the validation set\footnote{An extreme example, \emph{leave-one-out} cross validation, uses all but one observation as training data.} (often between 5:1 and 10:1 \cite{rasmussen2006gaussian}). 
    \item For a candidate set of hyperparameters, a GP is constructed using the training data, and used to estimate the mean and variance of the \emph{observations} in the validation set.
    \item The estimates are compared against the validation measurements, and a partial cost is formulated from the deviation.
    \item This process is repeated with different divisions of the data until all available measurements have been validated against. The total cost for this set of hyperparameters is calculated as sum over these batches.
\end{enumerate}

As with marginal likelihood maximisation, the cost function may have several local minima or may lack smoothness - the use of multi-start optimisation or a process such as simulated annealing can help avoid these \cite{hendriks2019robust}.

As a demonstration, we consider the use of $k-$fold cross validation on the example shown in Figure \ref{fig:fig1}.

Dividing the 200 observations into 10 bins, the cross validation approach determined a length scale nearly one order of magnitude smaller, subsequently allowing a better fit to the underlying function compared to marginal likelihood maximisation. The resulting reconstruction is shown in figure \ref{fig:fig2}.

\begin{figure}[h!]
\begin{center}
    \includegraphics[width=0.8\linewidth]{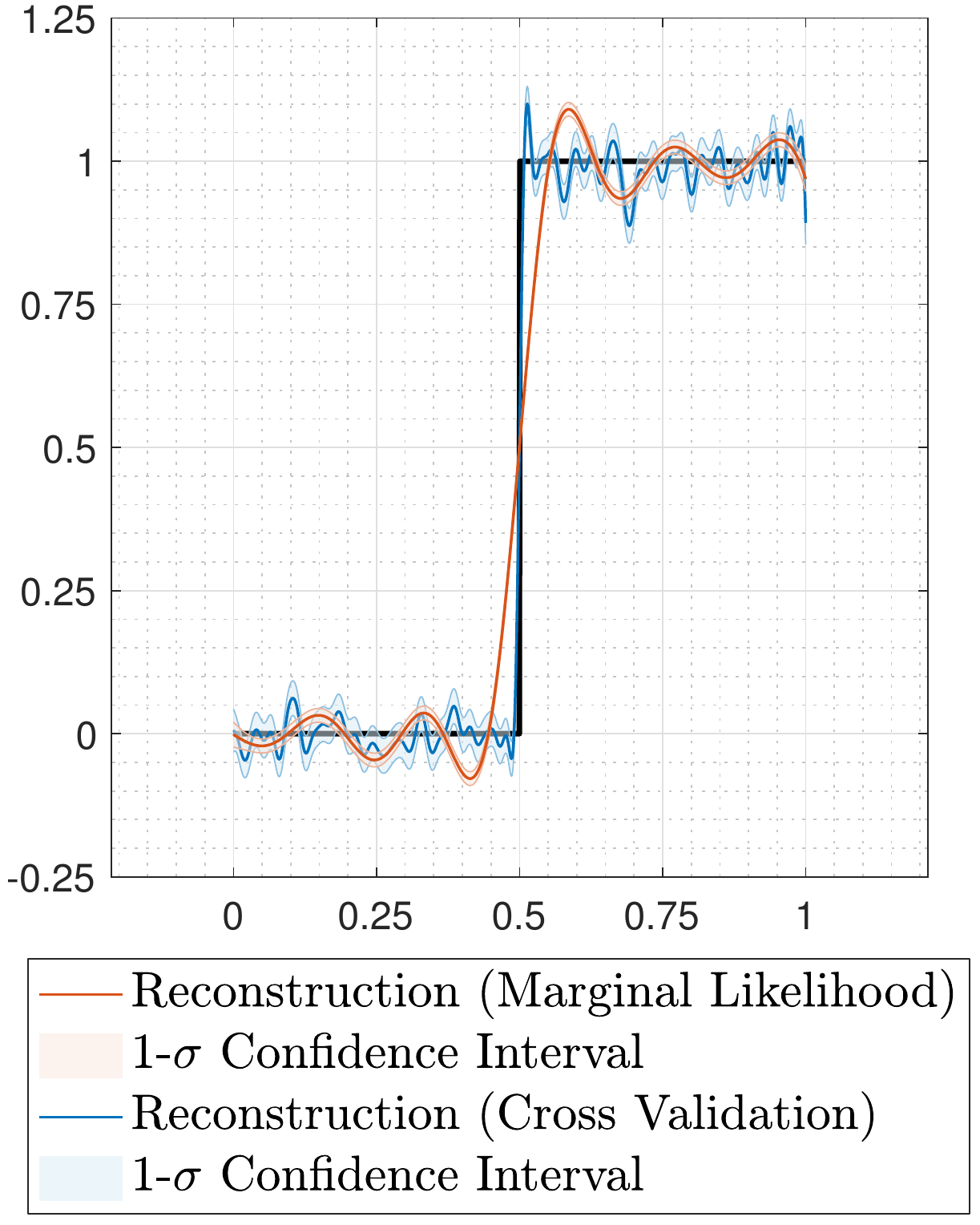}
\caption{Reconstructions from the simulated point-wise measurements shown in Figure \ref{fig:fig1} using two different hyperpameter selection processes.}
    \label{fig:fig2}
\end{center}
\end{figure} 
\FloatBarrier

While some overfit to the noisy measurements can be seen, quantitatively speaking, the cross-validation reconstruction was in-general twice as good as the maginal likelihood maximisation. These results are summaried in Table \ref{tab:crossvalvslikelihood}.

\begin{table}[h!]
\centering
\begin{tabular}{r|cc}
\multicolumn{1}{c|}{} & Marginal Likelihood & Cross-Validation \\
\hline
Length Scale $\ell$  & 0.09                & 0.0095           \\
Mean Abs Error       & 0.045               & 0.026            \\
RMS Error            & 0.09                & 0.049           
\end{tabular}
\caption{1D reconstruction results summary: Marginal likelihood maximiation vs Cross-validation.}
\label{tab:crossvalvslikelihood}
\end{table}

\FloatBarrier

\newpage
\subsection{Radial Basis Functions}
\label{subsec:RBF}

When implementing an approximation scheme, RBFs provide an alternative to the harmonic approach \cite{rasmussen2006gaussian}. These stationary functions are constructed such that they decay spatially and have negligible contribution to the reconstruction outside a well-defined region of influence.

In this work, we demonstrate the use of squared exponential RBFs, centred at $x=\mu_x$:
\begin{align*}
\begin{split}
    \phi_j(x) &= \text{exp}\left(\frac{-\sqrt{(x-\mu_x)^2}}{2\ell^2}\right), \\
    \Sigma_{p,jj} &= \sigma_f^2.
\end{split}
\end{align*}
Where $\sigma_f$ is a prior variance, and $l$ a length scale.

These functions have a number of potential benefits over their harmonic counterparts in the context of reconstructing discontinuous or high gradient fields:

\begin{enumerate}
    \item With a finite influence, the concentration of RBFs can be increased as needed to capture fine details (such as a step change) while maintaining a low `resolution' in areas where the function varies slowly to reduce computational burden and overfitting.
    \item In a sense, initial placement of RBFs is reasonably straightforward - there is no motivation to include any which are centred outside the sample, and the minimum density of functions can be calculated from the length scales by an interative approach.
\end{enumerate}

Note that in general, any spatially decaying function can be chosen as an RBF. Some investigation into Exponential and Matèrn \cite{rasmussen2006gaussian} basis functions was also conducted, though neither proved ideal for implementation of physical constraints or integral-based measurement models\footnote{The former having a discontinuity after differentiation and the latter lacking a convenient closed-form line integral for LRT measurements.}. Investigation into other potential RBFs has been identified as one avenue for future research.

That being said, with a limited region of influence and the improvements in hyperparameter optimisation provided by \kfold cross validation, the difficulties associated with the squared exponential were found to be sufficiently mitigated, while the benefits provided by it's simplicity for implementation of physical contraints were maintained.

Figure \ref{fig:fig3} compares reconstructions with harmonic and radial basis functions. Ringing artefacts are present in the harmonic reconstruction, while the effect of the discontinuity is spatially limited when using RBFs. That being said, some overfit is visible in the RBF reconstruction due to the small length scale and inherent flexibility of this model.

\begin{figure}[h!]
\begin{center}
    \includegraphics[width=0.8\linewidth]{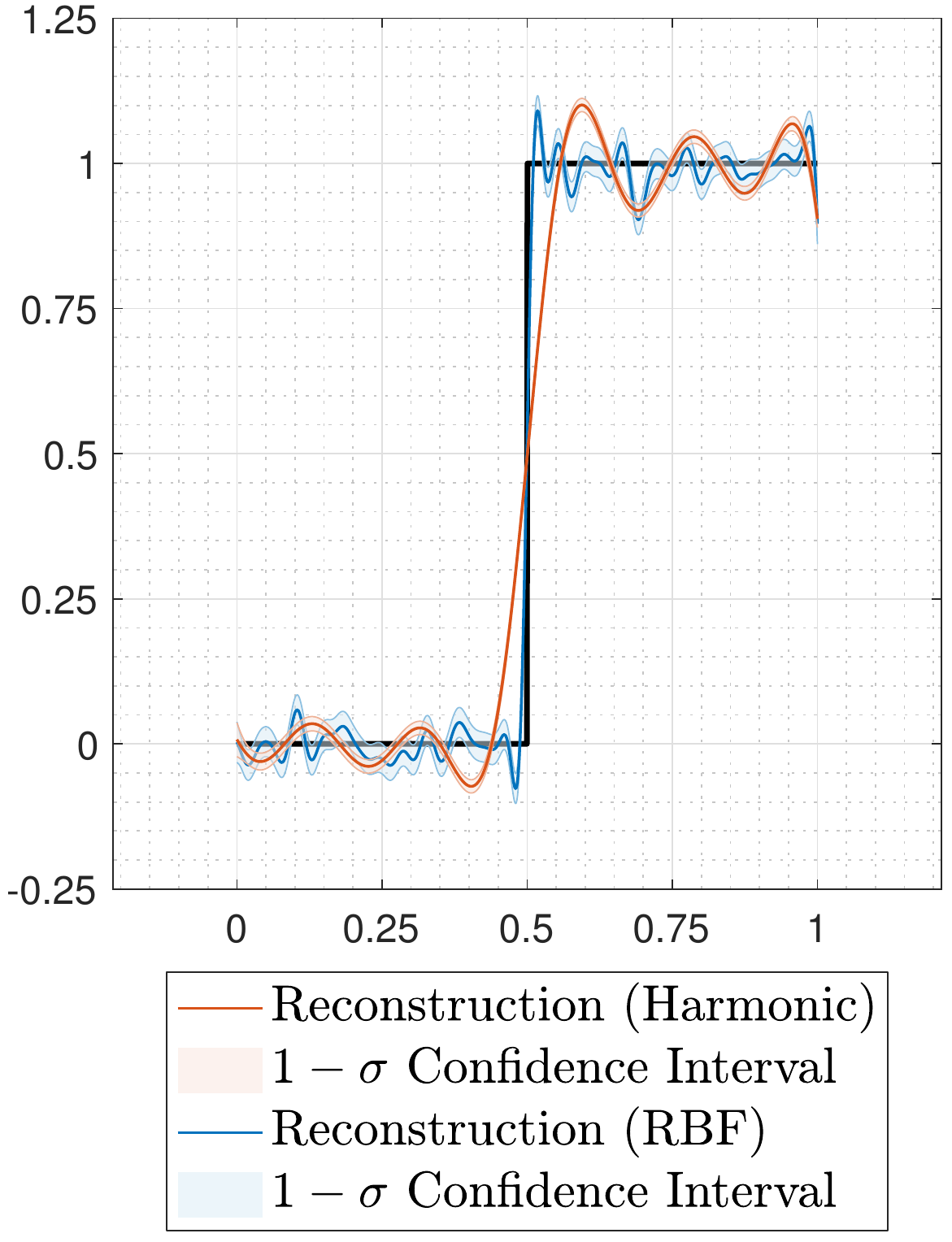}
\caption{Reconstructions using approximation schemes from the simulated pointwise measurements shown in Figure \ref{fig:fig1}.}
    \label{fig:fig3}
\end{center}
\end{figure} 
\FloatBarrier

Quantitative results are summaried in Table \ref{tab:harmonicvsrbf}. A significant improvement is seen by implementing the RBF approximation.

\begin{table}[h!]
\centering
\begin{tabular}{r|cc}
\multicolumn{1}{c|}{} & Harmonic & RBF \\
\hline
Mean Abs Error       & 0.06               & 0.034            \\
RMS Error            & 0.1                & 0.054           
\end{tabular}
\caption{1D reconstruction results summary: Harmonic vs Radial Basis Functions.}
\label{tab:harmonicvsrbf}
\end{table}

\FloatBarrier

\subsection{Gradient-based Basis Function Placement}
\label{subsec:gradient}

The usefulness of RBFs is particularly evident when exploiting the ability to adjust their density and placement as needed.

The freedom to independently control the length scale of individual or groups of RBFs also allows for spatial variation in $\ell$ that has the potential to provide both resolution in areas of high gradient and a reduction in overfitting in areas of low gradient.

To this end, we propose a rudimentary algorithm for gradient-dependent RBF placement: 

\begin{enumerate}
    \item Distribute a dense initial set of equally spaced RBFs on the sample domain. The minimum spacing can be calculated from the optimised length scales by an interative approach.
    \item Reconstruct, using cross-validation for hyperparameter optimisation.
    \item Calculate the gradient of the reconstruction. This can be achieved with numerical derivatives, or the GP can include the derivative as another quantity to estimate.
    \item Place a second set of RBFs in areas of high gradient, with the first and second sets having independent length scales.
    \item Reconstruct again. Typically, the hyperparameter optimisation will increase the length scale in areas of low gradient, reducing overfit and providing a smother solution, while, due to a limited region of influence, the length scales of RBFs in regions of high gradient is typically increased, and a better fit is obtained.
\end{enumerate}

This is certainly not the optimal placement method, and possible improvements to this approach are discussed in Section \ref{sec:conclusion}.

We again demonstrate this concept on the 1-D unit step example. As shown in Figure \ref{fig:fig4}, an initial reconstruction is obtained from an equally-spaced primary set of RBFs. The gradient of the reconstruction is then found, and by thresholding, a region of high gradient is identified (here, a 30\% threshold was used). A refinement set of RBFs is then placed in this region and the reconstruction re-run. As anticipated, the hyperparameter optimisation relaxes the length scale in the areas of low gradient and tightens the length scale in the region of high gradient. The result is a closer fit to the discontinuity, and a reduction in overfit of noise.

Quantitative results are summarised in Table \ref{tab:1vs2stage}.

\begin{table}[h!]
\centering
\begin{tabular}{r|cc}
\multicolumn{1}{c|}{} & 1-stage & 2-stage \\
\hline
Length Scale $\ell$  & 0.023                & $0.055,0.009$           \\
Mean Abs Error       & 0.034               & 0.021            \\
RMS Error            & 0.054                & 0.039           
\end{tabular}
\caption{1D reconstruction results summary: 1-stage vs 2-stage RBF reconstructions.}
\label{tab:1vs2stage}
\end{table}

\FloatBarrier

\begin{figure}[h!]
\begin{center}
    \includegraphics[width=0.8\linewidth]{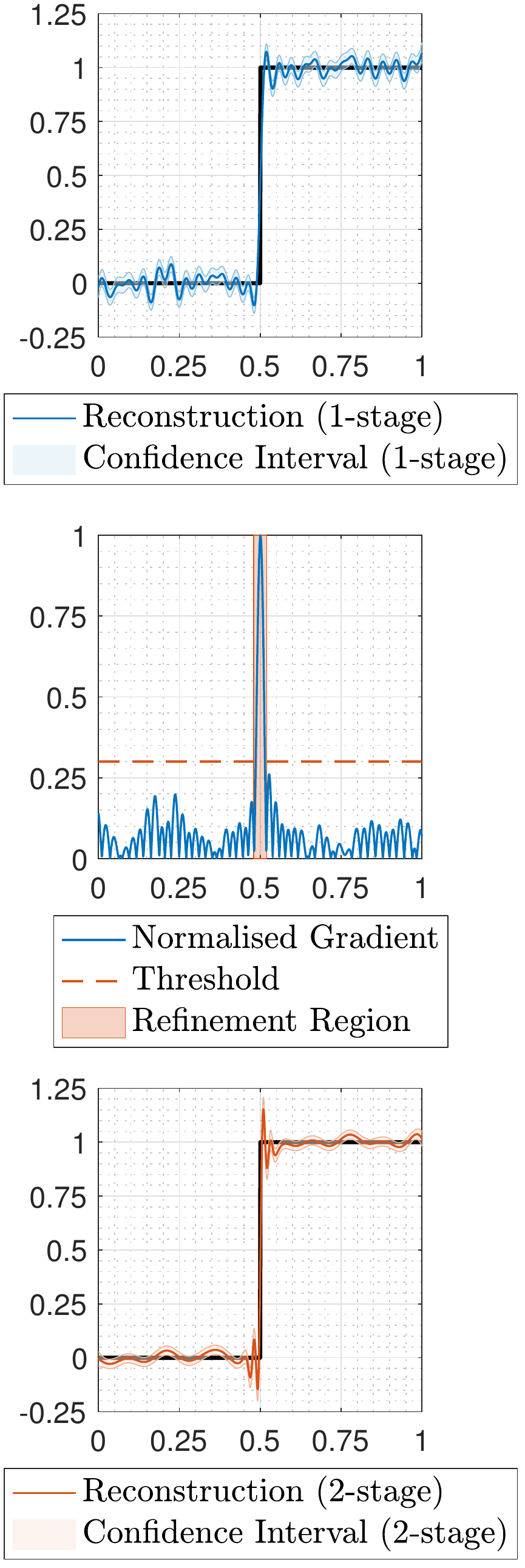}
    \caption{Top: First-stage reconstruction using 100 equally spaced radial basis functions. Centre: Reconstruction gradient (normalised) and identified refinement region. Bottom: Second-stage reconstruction.}
    \label{fig:fig4}
\end{center}
\end{figure} 

\FloatBarrier
\newpage
\section{Demonstration: Bragg-edge Neutron Strain Tomography}
\label{sec:exp}
To demonstrate these techniques in the context of strain reconstruction, a modified version of the algorithm presented in \cite{jidling2018probabilistic} was constructed using RBFs, \kfold cross validation for hyperparameter optimisation, and a two-stage gradient-based RBF placement method. Reconstructions are compared against the unmodified algorithm originally published in \cite{jidling2018probabilistic}, which uses a harmonic approximation and marginal likelihood maximisation for hyperparameter optimisation.

The three in-plane components of strain within a two dimensional (plane stress) sample were reconstructed from a Bragg-edge neutron transmission measurement set collected during an experiment at the Japan Proton Accelerator Research Complex (J-PARC) in 2018. The through-thickness average strain measurements were from a small, EN-26 steel offset ring-and-plug shrink-fit sample that exhibits a discontinuity in strain. Details concerning the experiment, sample and measurement pre-processing can be found in \cite{gregg2018resid} and \cite{hendriks2018traction}.

To summarise, 50 projections, each with a sampling time of 2 hours at a source power of 409 kW were obtained. Neutron counts were binned over columns of the detector to provide 1-D profiles of strain. In this work, neutron counts were binned over 5 column increments to provide an average uncertainty around $0.6\times10^{-4}$. This binning provided around 90 measurements per projection for a total of 4500 LRT observations.

200 synthetic free-stress traction measurements as described in \cite{hendriks2018traction} were also evenly distributed over the boundary of the sample.

Reconstructions are shown in Figure \ref{fig:fig6} where they are validated against conventional diffraction strain scans performed on the KOWARI diffractometer within ANSTO. The details of this validation experiment are also provided in \cite{gregg2018resid}. To summarise: measurements of the three in-plane components of strain at 195 points within the sample were performed based on the relative shift of the (211) diffraction peak using an $0.5\times0.5\times14$mm$^3$ gauge volume. The measurement locations are shown in Figure \ref{fig:fig6}, as well as an interpolated strain map. Note that this map has been constructed with separate interpolants for the ring and plug --- an appropriate use of a-priori knowledge given that this serves as a reference against which to compare our reconstructions.

The harmonic reconstruction was conducted using 7750 basis functions with frequencies determined according to the spectral density of the measurements and hyperparameters tuned using marginal likehihood maximisation, --- $\ell_x = 3.5$mm and $\ell_y = 4.1$mm. To provide indicative quantitative results, the mean absolute and RMS difference between the estimated strains and KOWARI meausurements are summaried in Table \ref{tab:mainreconresults}.

The first-stage of the RBF reconstruction was conducted using a primary set of 7668 equally spaced squared-exponential RBFs (nominally $0.25\times0.25$mm spacing between centres) within the boundary of the sample. Hyperparameters were tuned using \kfold cross validation --- $\ell_{x1} = 1.55$mm and $\ell_{y1} = 1.57$mm --- and applied to all RBFs in the primary set. This first-stage reconstruction is also shown in Figure \ref{fig:fig6} with quantitative results again summaried in Table \ref{tab:mainreconresults}.

Following the first-stage reconstruction, a norm of the 6 pertinent directional derivatives of strain was calculated as follows:

\begin{align*}
G &= \frac{F}{\max (F)}, \\
F & = \norm{\begin{bmatrix}\frac{\partial \epsilon_{xx}}{\partial x} & \frac{\partial \epsilon_{xx}}{\partial y} & \frac{\partial \epsilon_{xy}}{\partial x} & \frac{\partial \epsilon_{xy}}{\partial y} & \frac{\partial \epsilon_{yy}}{\partial x} & \frac{\partial \epsilon_{yy}}{\partial y} \end{bmatrix}}_2.
\end{align*}

This distribution is shown in Figure \ref{fig:fig5}. Also shown are identified regions of high gradient using simple thresholding. The technique was able to identify the high gradient region around the ring-plug boundary with no a-priori knowledge.

\begin{figure}[h!]
\begin{center}
    \includegraphics[width=0.8\linewidth]{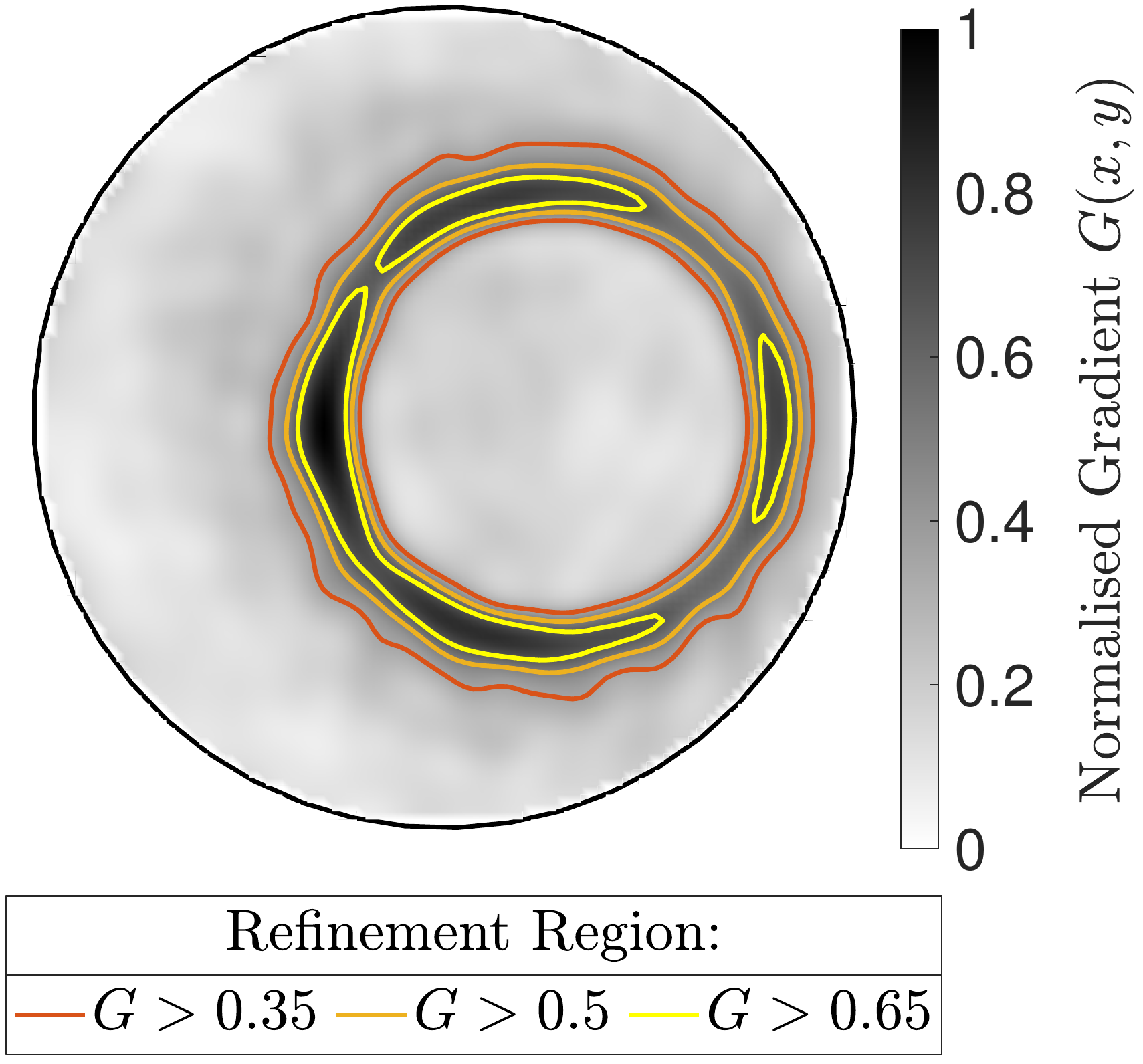}
    \caption{Normalised gradient of the 1st-stage RBF reconstruction and high-gradient areas identified by thresholding.}
    \label{fig:fig5}
\end{center}
\end{figure} 

With the high gradient region identified (using a 50\% threshold), a second-stage reconstruction was conducted. A refinement set of RBFs (with hyperparameters $\ell_{x2}$ and $\ell_{y2}$ independent of the first) was placed on an equally-spaced grid within the identified region of high gradient at a 0.125mm resolution. As in the first stage, \kfold cross validation was used to find hyperparameters. As previously seen in the 1D example, this process automatically relaxed the length scales of the primary set and tightened that of the refinement set: $\ell_{x1} =3$mm, $\ell_{y1} =2.4$mm, $\ell_{x2} = 1.4$mm $\ell_{y2} =0.95$mm. These reconstruction results are also shown in Figure \ref{fig:fig6} and quantitatively summaried in Table \ref{tab:mainreconresults}.

\begin{table}[h!]
\centering
\begin{tabular}{r|ccc}
\multicolumn{1}{c|}{} & Harmonic & RBF: 1-stage & 2-stage \\
\hline
Mean Abs Difference       & $122\pm16\mu\epsilon$               & $114\pm18\mu\epsilon$            & $106\pm17\mu\epsilon$            \\
RMS Difference            & $158\pm9\mu\epsilon$                & $145\pm8\mu\epsilon$           & $134\pm6\mu\epsilon$            
\end{tabular}
\caption{Primary Reconstruction Results. For this data set, the proposed developments to the GP method provide modest improvements compared to the previous approach.}
\label{tab:mainreconresults}
\end{table}

As expected, the reconstruction from the harmonic approximation using marginal likelihood for hyperparameter optimisation is notably smoother than that using RBFs and \kfold cross validation. That being said, the developments to the GP technique we propose allow a marginally better fit - particularly near the discontinuity. As summaried in Table \ref{tab:mainreconresults}, a small reduction in both mean absolute and RMS difference compared to the KOWARI measurements is noted with the new developments. Note of course that these measurements are not a ground truth and have their own uncertainty.

While positive, these results are not entirely indicative of the potential gains from the developments we propose due to the high level of noise in the measured data and the lack of a ground truth. In \ref{app:sim}, results from a simulated measurement set are discussed and the potential improvements of the proposed approach are clearer.

\section{Conclusion and Future Work}
\label{sec:conclusion}

In this proof-of-concept study, we provided three improvements to the GP technique to improve performance primarily for strain fields exhibiting high gradients or discontinuities in strain. A combination of \kfold Cross-Validation for selecting hyperparameters, and Gradient-based placement of Radial Basis Functions was able to obtain a substantially better reconstruction of a discontinuous strain field from experimental Bragg-edge neutron transmission measurements compared to the prior published approach using marginal likelihood and harmonic basis functions.

Many further improvements to these techniques can be made, and investigation of the following forms a natural basis for future work;
\begin{enumerate}
    \item Other possible choices of RBF may prove more suitable but were not investigated. Nominally any stationary function could be used as an RBF and a better alternative may yet be found.
    \item In this work, a simple threshold was used on the calculated gradient to determine where to place a refinement set of RBFs. A number of alternatives may yield better results, including:
    \begin{enumerate}
    \item Intelligent meshing of basis functions based on the gradient - some inspiration from finite-element approaches may be useful here.
    \item Further iteration in the multi-stage approach to place 3, 4 or more sets of RBFs. With each set the hyperparameter optimisation becomes more challenging, but this additional computational burden may be eased slightly as resolution far from the discontinuity could be significantly reduced. 
    \item Taking the previous point to the extreme --- individual length scales for each RBF, and/or optimising the position of each RBF. This would be challenging as it would substantially increase the number of hyperparameters. Potential tools to solve this problem could come from areas such as machine learning, where neural networks with tens of thousands of parameters are trained using e.g. stochastic gradient descent \cite{kingma2014adam}.
    \item Use of the calculated gradient to determine not only the position of basis functions, but to help inform the required length scales in combination with cross-validation. One potential approach would be to use cross validation to determine the parameters of a gradient-dependent function that defines a distribution of the length scales over the sample geometry. 
    \end{enumerate}
\end{enumerate}

\section{Acknowledgements}
This work is supported by the Australian Research Council through a Discovery Project Grant (DP170102324). Access to the RADEN and KOWARI instruments was made possible through the respective user access programs of J-PARC and ANSTO (J-PARC Long Term Proposal 2017L0101 and ANSTO Program Proposal PP6050). The authors would also like to thank AINSE Limited for providing financial assistance (PGRA) and support to enable work on this project.

\newpage
\begin{figure*}[h!]
\begin{center}
    \includegraphics[width=\linewidth]{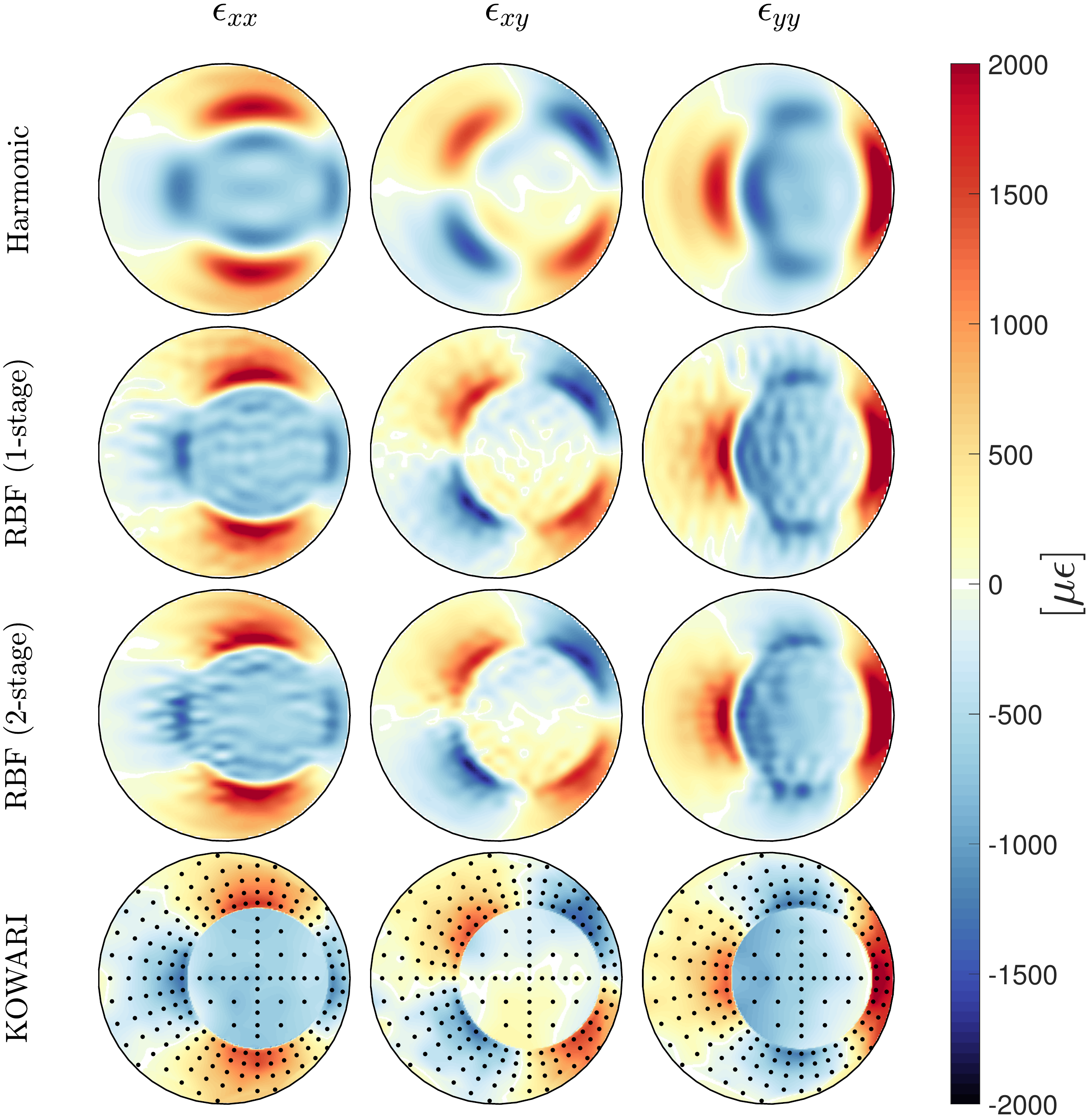}
    \caption{Reconstruction results from the experimental data set. Qualitatively and quantitatively, the 2-stage RBF reconstruction using \kfold cross validation outperforms the previous method.}
    \label{fig:fig6}
\end{center}
\end{figure*} 
\FloatBarrier

\newpage
\bibliography{references}

\appendix

\section{Equilibrium Constraints for Radial Basis Functions}

In this paper we implement squared-exponential radial basis functions of the form:

\begin{equation*}
\phi_j(\boldsymbol{x}) = \text{exp}\left(\frac{-(x-\mu_x)^2}{2\ell_x^2} \frac{-(y-\mu_y)^2}{2\ell_y^2}\right).
\end{equation*}

To encode equilibrium, we construct these basis functions to represent an Airy stress function. In 2D and assuming plane stress, these are then related to the components of strain by:

\begin{equation*}
\boldsymbol{\epsilon} = \begin{bmatrix} \epsilon_{xx} \\ \epsilon_{xy} \\ \epsilon_{yy}\end{bmatrix} = \begin{bmatrix} \frac{\partial^2}{\partial y^2}- \nu\frac{\partial^2}{\partial x^2}\\ -(1+\nu)\frac{\partial^2}{\partial x \partial y} \\ \frac{\partial^2}{\partial x^2}- \nu\frac{\partial^2}{\partial y^2}\end{bmatrix} \phi_j(\boldsymbol{x}).
\end{equation*}

This means a linear combination of second derivates of the basis function are required for reconstruction. These have a closed form as follows:

\begin{align*}
\frac{\partial^2}{\partial x^2} &= \frac{(x-\mu_x)^2-\ell_x^2}{\ell_x^4}\phi_j(\boldsymbol{x}) \\
\frac{\partial^2}{\partial y^2} &= \frac{(y-\mu_y)^2-\ell_y^2}{\ell_y^4}\phi_j(\boldsymbol{x})\\
\frac{\partial^2}{\partial x \partial y} &= \frac{(x-\mu_x)(y-\mu_y)}{\ell_x^2\ell_y^2}\phi_j(\boldsymbol{x})
\end{align*}

And thus the strains can be written in terms of the basis functions by:

\begin{align*}
\epsilon_{xx} &= \phi_j(\boldsymbol{x}) \left( \frac{(y-\mu_y)^2-\ell_y^2}{\ell_y^4} -\nu \frac{(x-\mu_x)^2-\ell_x^2}{\ell_x^4} \right) \\
\epsilon_{xy} &= \phi_j(\boldsymbol{x}) \left( -(1+\nu) \frac{(x-\mu_x)(y-\mu_y)}{\ell_x^2\ell_y^2}\right) \\
\epsilon_{yy} &= \phi_j(\boldsymbol{x}) \left(\frac{(x-\mu_x)^2-\ell_x^2}{\ell_x^4} -\nu\frac{(y-\mu_y)^2-\ell_y^2}{\ell_y^4}  \right)
\end{align*}

\section{Longitudinal Ray Transform of Radial Basis Functions (for Bragg-edge Neutron Transmission Measurements)}

The measurement model for Bragg-edge neutron transmission measurements is the longitudinal ray transform, which, with respect the sample geometry and co-ordinate system in \cite{gregg2018resid}, is given by:
\begin{equation*}
y=\frac{1}{L}\int_0^L \nhat^\Transp \bepsilon (s) \nhat  \dd s
\end{equation*}

Applying linearity and expanding, we have:

\begin{equation*}
y = \frac{n_x^2}{L}\int_0^L  \epsilon_{xx}(s) \dd s + \frac{2n_xn_y}{L}\int_0^L \epsilon_{xy}(s) \dd s + \frac{n_y^2}{L}\int_0^L \epsilon_{yy}(s) \dd s
\end{equation*}

After substituting the previously determined expressions for $\epsilon_{xx}$, $\epsilon_{xy}$ and $\epsilon_{yy}$ in terms of $\phi_j$, and making the co-ordinate transformations $x = x_0 + s n_x$ and $x = y_0 + s n_y$, where $x_0$ and $y_0$ are the entry co-ordinates of the ray and sample, line integrals of each component of strain must be performed. These line integrals have a closed form:
\begin{align*}
&\int_0^L  \epsilon_{ij}(s) \dd s = G\bigg(\frac{C_{ij} + B E_{ij}}{4\sqrt{A}^3} -\frac{B^{2} C_{ij}}{8\sqrt{A}^5 }-\frac{F_{ij}}{2\sqrt{A}}\bigg) \\
& \qquad -\frac{(\phi_0-D)(2AE_{ij}+B C_{ij}) + 2AC_{ij} D L}{4A^2} \\
\end{align*}
Where:
\begin{align*}
&\delta_{i0} = i_0-\mu_i, \qquad \phi_0 = \exp\left(\frac{-(\ell_y^2\delta_{x0}^2+\ell_x^2\delta_{y0}^2)}{2\ell_x^2\ell_y^2}\right) \\
&A = \frac{-(\ell_y^2n_x^2 + \ell_x^2n_y^2)}{2\ell_x^2\ell_y^2}, B = \frac{-(n_x\ell_y^2\delta_{x0}+n_y\ell_x^2\delta_{y0})}{\ell_x^2\ell_y^2} \\
&C_{ij} = \frac{n_i n_j}{\ell_i^2\ell_j^2}, D = \phi_0\exp\left(BL + AL^2\right), E_{ij} = \frac{n_j\delta_{i0} + n_i\delta_{j0}}{\ell_i^2\ell_j^2} \\
&F_{xx} = \frac{\delta_{x0}^2-\ell_x^2}{\ell_x^4}, F_{xy} = \frac{\delta_{x0}\delta_{y0}}{\ell_x^2\ell_y^2}, F_{yy} = \frac{\delta_{y0}^2-\ell_y^2}{\ell_y^4} \\
&G = \phi_0\sqrt{\pi}\exp\left(-\frac{B^2}{4A}\right) \bigg(\text{erfi}\left(\frac{B}{2\sqrt{A}}\right) - \text{erfi}\left(\frac{B+2AL}{2\sqrt{A}}\right)\bigg) \\
\end{align*}

\newpage
\section{Simulation Results}
\label{app:sim}

While positive, the results presented in Section \ref{sec:exp} are not entirely indicative of the potential of the proposed developments. To illustrate this point, a reconstruction from simulated LRT measurements of a finite-element model of the ring-and-plug strain field was performed.

A measurement set maintaining the same number and distribution of rays  as the experimental data (i.e. 4500 measurements over 50 projection angles) was constructed by applying the LRT to the finite-element model. The distribution of noise within these measurement was also maintained, but halved to an average standard deviation of $0.25\times10^{-4}$.

Note that while, generally speaking, to achieve this of confidence would require quadruple the sampling time, this quality of data may well be achieveable in the future with sources continually increasing in brightness --- J-PARC, for example now operates at 600 kW compared to the 409 when this experiment was conducted and is projected to reach 1MW in near future).

Reconstruction results are shown in Figure \ref{fig:fig7} and are summarised in Table \ref{tab:appendixreconresults}. Notably, the new approach benefits most from these higher quality measurements, while --- limited by the smooth modelling assumption inherent to both the approximation and marginal likelihood maximisation process --- the previous approach \cite{hendriks2018traction} only demonstrates a minor improvement compared to the noisier experimental data.

\begin{table}[h!]
\centering
\begin{tabular}{r|ccc}
\multicolumn{1}{c|}{} & Harmonic & RBF: 1-stage & 2-stage \\
\hline
Mean Abs Error       & 103$\mu\epsilon$               & 75$\mu\epsilon$            & 69$\mu\epsilon$            \\
RMS Error            & 198$\mu\epsilon$                & 129$\mu\epsilon$           & 114$\mu\epsilon$            
\end{tabular}
\caption{Simulated Reconstruction Results.}
\label{tab:appendixreconresults}
\end{table}

\begin{figure*}[h!]
\begin{center}
    \includegraphics[width=\linewidth]{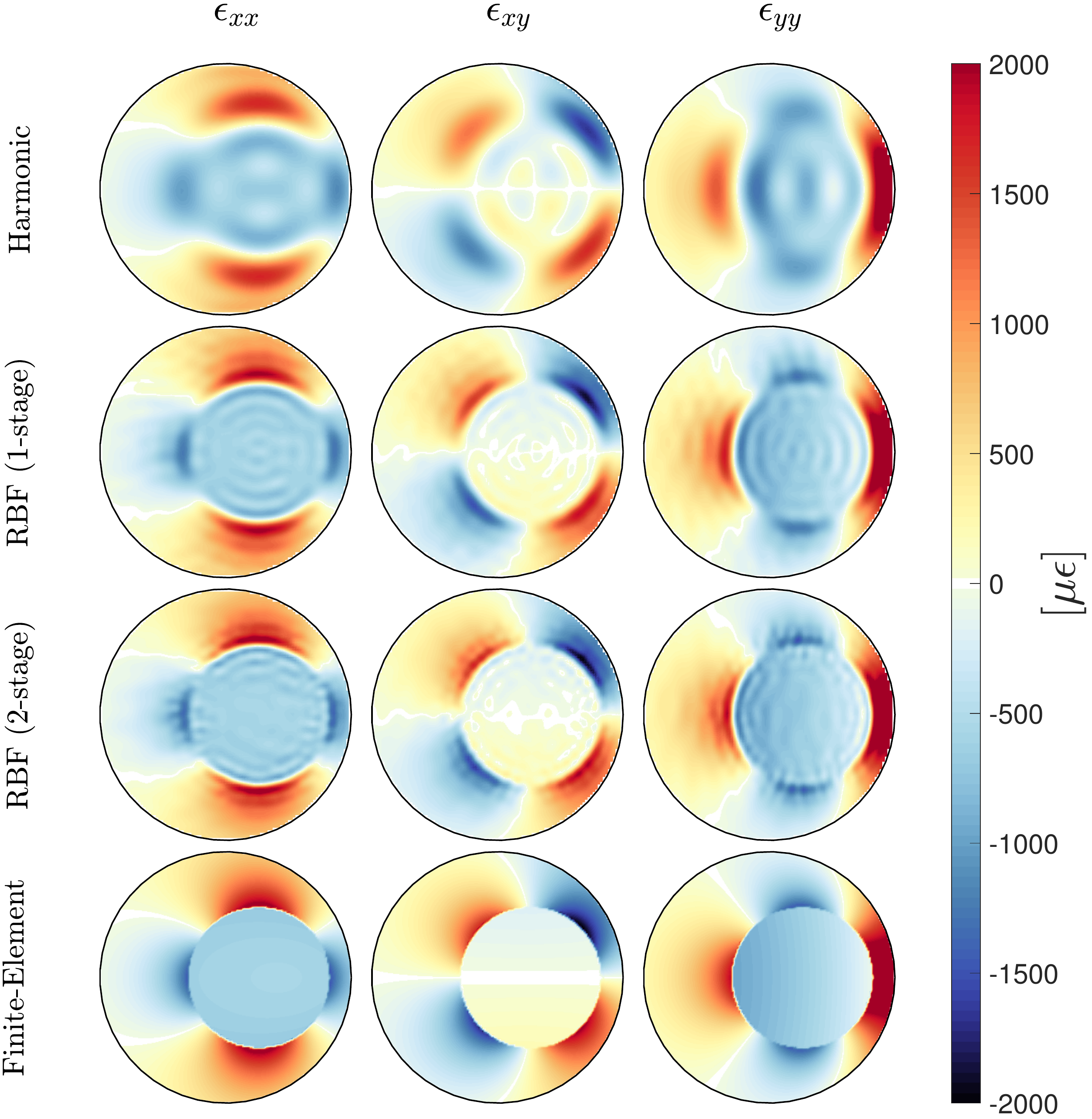}
    \caption{Reconstruction results from a simulated measurement set. The 2-stage RBF reconstruction using \kfold cross validation demonstrates superior performance compared to the previous method.}
    \label{fig:fig7}
\end{center}
\end{figure*} 

\end{document}